\documentclass{ws-procs9x6}

\begin{document}

\newcommand{\lsim}{\mbox{\raisebox{-.9ex}{~$\stackrel{\mbox{$<$}}{\sim}$~}}}
\newcommand{\gsim}{\mbox{\raisebox{-.9ex}{~$\stackrel{\mbox{$>$}}{\sim}$~}}}

\title{ TRAPPED QUINTESSENTIAL INFLATION\\
FROM FLUX COMPACTIFICATIONS}

\author{K. DIMOPOULOS$^*$}

\address{Physics Department, Lancaster University,\\
Lancaster LA1 4YB, U.K.\\
$^*$E-mail: k.dimopoulos1@lancaster.ac.uk}

\begin{abstract}
Quintessential inflation is studied using a string modulus as the inflaton--%
quintessence field. It is assumed that the modulus crosses an enhanced symmetry
point (ESP) in field space. Particle production at the ESP temporarily traps 
the modulus resulting in a period of inflation. After reheating, the modulus 
freezes due to cosmological friction at a large value, such that its scalar 
potential is dominated by contributions due to fluxes in the extra dimensions. 
The modulus remains frozen until the present, when it can become quintessence.
\end{abstract}


\bodymatter

\section{Introduction}\label{Int}
A plethora of observations concur that the Universe at present enters a
phase of accelerated expansion. In fact, most cosmologists accept that
over 70\% of the Universe content at present corresponds to the elusive
dark energy; a substance with pressure negative enough to cause 
the observed acceleration \cite{DE}. The simplest form of dark 
energy is a positive cosmological constant $\Lambda$, which however, needs to 
be incredibly fine-tuned to explain the observations \cite{L}. This is why 
theorists have looked for alternatives, which could explain the observations 
while setting \mbox{$\Lambda=0$}, as was originally assumed. A promising idea 
is to consider that the Universe at present is entering a late-time 
inflationary period \cite{early}. The credibility of this option is supported
also by the fact that the generic predictions of inflation in the 
early Universe are in excellent agreement with the observations. The scalar 
field responsible for this late-inflation period is called quintessence because
it is the fifth element after baryons, photons, CDM and neutrinos \cite{Q}. 

Since they are based on the same idea, it is natural to attempt to unify
early Universe inflation with quintessence. Quintessential inflation was thus 
born \cite{quinf,QI,jose,eta}. This attempt has many advantages. Firstly,
quintessential inflation models allow the treatment of both inflation and 
quintessence within a single theoretical framework. Also, quintessential 
inflation dispenses with the tuning problem of the initial 
conditions for quintessence. Finally, unified models for inflation and 
quintessence are more economic because they avoid introducing yet 
another unobserved scalar field.

For quintessential inflation to work one needs a scalar field with a runaway
potential, such that the minimum has not been reached until today and, 
therefore, there is residual potential density, which can cause 
the observed accelerated expansion. String moduli fields are suitable 
because they are typically characterised by such runaway potentials. 
The problem with such fields, however, is how to stabilise them temporarily, in
order to use them as inflatons in the early Universe. In this work 
(see also Ref.~\cite{ours}) we achieve this by considering that, during its 
early evolution our modulus crosses an enhanced symmetry point (ESP) in field 
space. When this occurs the modulus is trapped temporarily at the ESP 
\cite{trap}, which leads to a period of inflation. After inflation the modulus 
picks up speed again in field space resulting into a period of kinetic density 
domination (kination) \cite{kination}. Kination ends when the thermal 
bath of the hot big bang (HBB) takes over. During the HBB, due to cosmological 
friction \cite{cosmofric}, the modulus freezes at some large value and remains 
there until the present, when its potential density dominates and drives the 
late-time accelerated expansion \cite{eta}.

Is is evident that, in order for the modulus to become quintessence, it should 
not decay after the end of inflation. Reheating, therefore should be achieved 
by other means. We assume that the thermal bath of the HBB is due to the decay 
of some curvaton field \cite{curv} as suggested in Refs.~\cite{eta,curvreh}.
By considering a curvaton we do not add an {\it ad~hoc\/} degree of freedom,
because the curvaton can be a realistic field, already present in simple 
extensions of the standard model (e.g. a right-handed 
sneutrino \cite{sneu}, a flat direction of the (N)MSSM \cite{mssm}
or a pseudo Nambu-Goldstone boson \cite{pngb,orth} possibly associated with the
Peccei-Quinn symmetry \cite{PQ}). Apart from reheating, the curvaton can
provide the correct amplitude of curvature perturbations in the Universe.
Consequently, the energy scale of inflation can be much 
lower than the grand unified scale \cite{liber}. In fact, in certain curvaton 
models, the Hubble scale during inflation can be as low as the electroweak 
scale \cite{orth,low}.

\section{The runaway scalar potential}

String theories contain a number of flat directions which are parametrised by
the so-called moduli fields, which correspond to the size and shape of the 
compactified extra dimensions. Many such flat directions are lifted by
non-perturbative effects, such as gaugino condensation or D-brane instantons
\cite{Derendinger:1985kk}. The superpotential, then, is of the form
\begin{equation}
W=W_0+W_{\rm np}\quad \textrm{with} \quad W_{\rm np}=Ae^{-cT}\,,
\end{equation}
where $W_0\approx$~const. is the tree level contribution from fluxes, 
$A$ and $c$ are constants and $T$ is a K\"ahler modulus in units of $m_P$.
Hence, the non-perturbative superpotential $W_{\rm np}$
results in a runaway scalar potential characteristic of string
compactifications. For example, in type IIB compactifications with a single
K\"ahler modulus, $\sigma\equiv$~Re($T$) is the so-called volume modulus, which
parametrises the volume of the compactified space. In this case,
the runaway behaviour leads to decompactification of the internal manifold.
The tree level K\"ahler potential for a modulus, in units of $m_P^2$, is
\begin{equation}
  K=-3\,\ln\,(T+\bar{T})\equiv-3\ln(2\sigma)\,,
\label{tree}
\end{equation}
and the corresponding supergravity potential is%
\footnote{We considered $c\sigma>1$ to secure the validity of the supergravity 
approximation and we have assumed that the ESP lies at a minimum in the 
direction of Im($T$).}
\begin{equation}
\label{Vnp}
 V_{\rm np}(\sigma)\simeq
\frac{cAe^{-c\sigma}}{2\sigma^2m_P^2}
\left(\frac{c\sigma}{3}Ae^{-c\sigma}-W_0\right)\,.
\end{equation}

To study the cosmology, we turn to the canonically normalised 
modulus $\phi$ which, due to Eq.~(\ref{tree}),
is associated with $\sigma$ as
\begin{eqnarray}
\label{fs}
&
\sigma(\phi)=
\exp\left(\sqrt{\frac{2}{3}}\,
\phi/m_P\right)\,.
&
\end{eqnarray}

Suppose that the Universe is initially dominated by the above modulus.
The non-perturbative scalar potential in Eq.~(\ref{Vnp}) 
is very steep (exponential of an exponential), which means that the field
soon becomes dominated by its kinetic density. Once this is so, the 
particular form of the potential ceases to be of importance. 
To achieve inflation we assume that, while rolling, the 
modulus crosses an ESP and becomes temporarily trapped at~it.

\section{At the Enhanced Symmetry Point}

In string compactifications there are distinguished points in
moduli space at which there is enhancement of the gauge symmetries 
\cite{Hull:1995mz}. This results in some massive states of the 
theory becoming massless at these points. 

Even though from the classical point of view an ESP is not a special point, as 
the modulus approaches it certain states in the string spectrum become massless
\cite{Watson:2004aq}. In turn, these massless modes create an interaction 
potential that may drive the field back to the symmetry point. In that way a 
modulus can become trapped at an ESP \cite{trap}. The strength of the symmetry
point depends on the degree of enhancement of the symmetry.

Such modulus trapping can lead to a period of so-called `trapped inflation'
\cite{trap}, when the trapping is strong enough to make the kinetic density of
the modulus fall below the potential density at the ESP. However, it turns out 
that the number of e-foldings of trapped inflation cannot be very large. 
Therefore, with respect to cosmology, the main virtue of the ESPs relies on 
their ability to trap the field and hold it there, at least temporarily. 

\begin{figure}
\hspace{1cm}\psfig{file=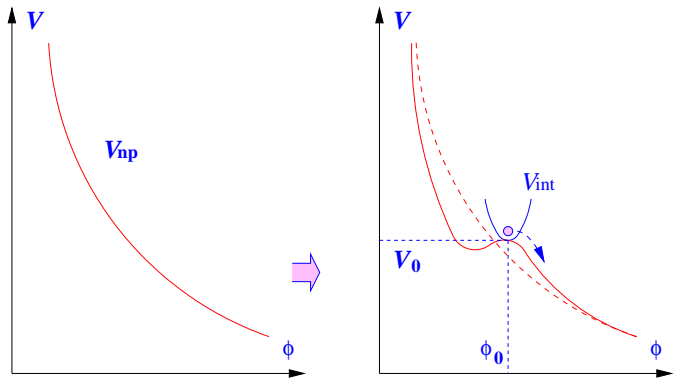,width=3.5in}
\caption{%
Illustration of how the appearance of an ESP at $\phi=\phi_0$ 
deforms the non-perturbative scalar potential $V_{\rm np}$ to generate, for 
example, a local maximum at potential density $V_0$. The crossing modulus is
temporarily trapped by the emergence of an interaction potential $V_{\rm int}$
due to its enhanced interaction with other fields. After released from 
trapping, the modulus may drive slow-roll inflation while sliding over the 
potential hill.}
\label{esp}
\end{figure}

Because ESPs are fixed points of the symmetries we have
\begin{equation}\label{eq:firstderiv}
V^{\prime}(\phi_0)=0\,,
\end{equation}
where the prime denotes derivative with respect to $\phi$ and
$\phi_0$ is the value of the modulus at the ESP.
The above means that the ESP is located either at a local extremum (maximum or 
minimum) or at a flat inflection point of the scalar potential, where
\mbox{$V'(\phi_0)=V''(\phi_0)=0$}.
This means that the presence of an ESP deforms the 
non-perturbative scalar potential (see Fig.~\ref{esp}). 
This deformation may be enough so that, after
trapped inflation, the field undergoes slow-roll inflation over the flat 
region of the scalar potential at the vicinity of the ESP. The total duration 
of inflation may, thus, be enough to solve the flatness and horizon problems 
of the HBB.

\subsection{Trapped Inflation}

Let us briefly study the trapping of the modulus at the ESP. 
We assume that around the ESP there is a contribution to the scalar potential
due to the enhanced interaction between the modulus $\phi$ and another field 
$\chi$, which we take to be also a scalar field \cite{trap}. The 
interaction potential is 
\begin{equation}
\label{Vint}
  V_{\rm int}(\phi,\chi)=\frac{1}{2}\,g^2\chi^2\bar\phi^2\,,
\end{equation}
where $\bar{\phi}\equiv\phi-\phi_0$ with 
$g$ being a dimensionless coupling constant.

Thus, at the ESP the $\chi$ particles are massless. The time
dependence of the effective (mass)$^2$ of the $\chi$ field results in the 
creation of $\chi$-particles.
This takes place when the field is within the production window 
$|\phi|<\Delta\phi\sim(\dot{\phi}_0/g)^{1/2}$,
where $\frac{1}{2}\dot\phi_0^2$ is the kinetic density of the modulus when
crossing the ESP and the dot denotes derivative with respect to the cosmic 
time~$t$. 

The effective scalar potential near the ESP is
$V_{\rm eff}(\phi)\approx V_0+\frac{1}{2}g^2\langle\chi^2\rangle\bar{\phi}^2$
where $V_0\equiv V(\phi_0)$ with $V(\phi)$ being the `background' scalar 
potential. Following Ref.~\cite{trap} we have 
$\langle\chi^2\rangle\simeq n_{\chi}/g|\phi|$, where $n_{\chi}$ denotes the 
number density of $\chi$ particles produced after the crossing of the ESP. 
This means that $V_{\rm eff}(\phi)\sim V_0+gn_{\chi}|\phi|$ and the field 
climbs a {\it linear} potential since $n_{\chi}$ is constant outside the 
production window. 

\begin{figure}
\hspace{2cm}\psfig{file=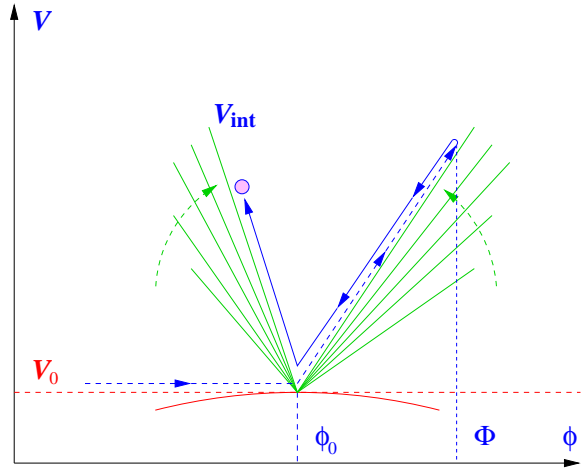,width=3in}
\caption{%
Illustration of the trapping of a modulus crossing the ESP during particle 
production. Outside the production window, the modulus oscillates in a linear 
interaction potential, which steepens progressively due to the production of 
more $\chi$-particles every time the modulus crosses the ESP.}
\label{trap1}
\end{figure}

After the first crossing, the field reaches the amplitude 
$\Phi_1$, determined by its initial kinetic density. To avoid overshooting the 
ESP we require $\Phi_1\lsim m_P$ since for 
larger values the coupling softens \cite{Brustein:2002mp}.
After reaching $\Phi_1$, the field reverses direction and crosses the 
production
window again, generating more $\chi$ particles and, therefore, increasing 
$n_\chi$. Thus, it now has to climb a steeper potential reaching an amplitude
$\Phi_2<\Phi_1$. The process continues until the ever decreasing amplitude 
becomes comparable to the production window (see Fig.~\ref{trap1}). At 
this moment particle production stops. 

\begin{figure}
\hspace{2cm}\psfig{file=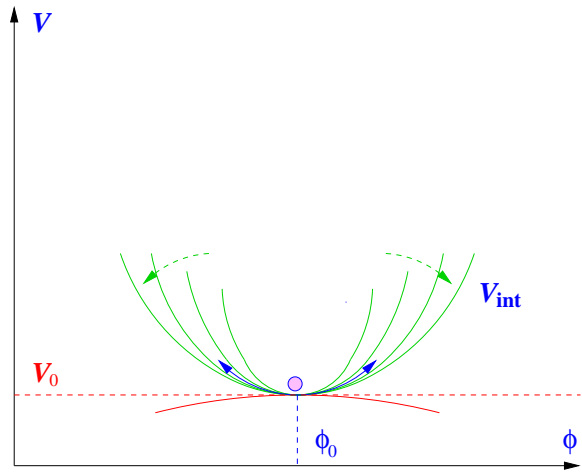,width=3in}
\caption{%
Illustration of the trapping of a modulus crossing the ESP after particle 
production. Inside the production window, the modulus oscillates in a quadratic
interaction potential, which becomes gradually diluted due to the Universe
expansion.}
\label{trap2}
\end{figure}

After the end of particle production, $\langle\chi^2\rangle$ remains roughly 
constant during an oscillation and the modulus continues oscillating in the 
quadratic interaction potential. Studying this oscillation,
we found that, due to the Universe expansion, the amplitude and frequency 
decrease as $\Phi\sim\Delta\phi/a$ and 
$\langle\overline{\chi^2}\rangle\propto a^{-2}$ \cite{ours},
where the scale factor $a(t)$ is normalised to unity at the end of 
particle production. Hence, the quadratic potential becomes gradually 
``diluted'' due to the Universe expansion (see Fig.~\ref{trap2}). The above 
mean that the kinetic density of the oscillating modulus scales as 
\mbox{$\rho_{\rm osc}\propto a^{-4}$}. When $\rho_{\rm osc}$ becomes redshifted
below $V_0$, trapped inflation begins. 

The above process assimilates a multitude of initial conditions (provided 
overshooting the ESP is avoided) because any kinetic density 
in excess of $V_0$ is depleted before the onset of trapped inflation.
Trapped inflation dilutes exponentially the density of the $\chi$--particles, 
which quickly redshifts $V_{\rm int}$. Therefore, after a rather limited number
of e-foldings of trapped inflation, the modulus is released from the ESP.

\subsection{Slow-Roll inflation}

Since the ESP is located at a locally flat region of the potential there
is a chance that, after $V_{\rm int}$ becomes negligible, the modulus drives a
period of slow-roll inflation while sliding away from the ESP. To study 
this period we need to quantify the deformation of the
scalar potential due to an ESP.

The appearance of an ESP generates either a local extremum or a flat 
inflection point at $\phi_0$. In all cases, in the vicinity of the ESP, the 
scalar potential can be approximated by a cubic polynomial\cite{ours}. 
Hence, the characteristics of the potential depend only on 
$m_\phi^2\equiv V''(\phi_0)$ and
$V^{\textrm{\tiny(3)}}_0\equiv V'''(\phi_0)$. In fact, we can parametrise the 
deformation of the scalar potential using \cite{ours}
\begin{equation}
\label{xi0}
|V^{\textrm{\tiny{(3)}}}_0|\sim
\xi^2\sigma_0^3H_*^2/m_P\,,
\end{equation}
where $\sigma_0\equiv\sigma(\phi_0)$ and $H_*$ is the Hubble parameter during 
inflation: $H_*^2\approx V_0/3m_P^2$. 
The $\xi$ parameter accounts for the strength of the symmetry point; the 
smaller the $\xi$, the stronger the deformation and 
the wider the inflationary plateau.
The requirement that the deformation becomes negligible at 
distances larger than $m_P$ results in the lower bound
$\xi>1$, which also guarantees that the modulus does not overshoot the ESP 
\cite{ours}.

By studying inflation after the modulus escapes trapping, we have obtained the 
following results, depending on the ESP morphology \cite{ours}. In each
case, one has to achieve enough inflationary e-folds to solve the horizon
and flatness problems, while also taking care that the curvature 
perturbations due to the modulus are not excessive compared to observations.

Consider first the case of a flat inflection point.
In this case, we can have enough e-foldings of slow roll inflation if
$|V^{\textrm{\tiny{(3)}}}_0|<g^2H_*\ll H_*$. 
The case of a local minimum is indistinguishable from the above if 
$m_\phi^2<g^2H_*|V^{\textrm{\tiny{(3)}}}_0|$. If the opposite is true then
the modulus becomes trapped in the local minimum and must
escape through tunnelling. Afterwards the modulus
can drive a period of slow-roll inflation with total number of e-foldings
given by $N\sim(H_*/m_\phi)^2$. Hence, to solve the horizon and flatness 
problems we need $m_\phi\ll H_*$. 
Finally, in the case of a local maximum,
after the end of trapping, one can have a phase of fast/slow roll
inflation provided $|m_\phi|\lsim H_*$.

Thus, we have found that, in all cases, {\em enough slow-roll 
inflation to solve the horizon and flatness problems of the HBB is attainable
provided} $|m_\phi|, |V^{\textrm{\tiny{(3)}}}_0|<H_*$. 
Choosing for illustrative purposes an intermediate value for the Hubble scale: 
$H_*\sim 1$~TeV we have found that one can achieve enough inflationary 
e-foldings (up to $N_{\rm max}\sim 10^4$)
without producing excessive curvature perturbation if $1<\xi^2<10^4$. Thus, 
{\em there is ample parameter space for slow-roll inflation to occur 
after the modulus escapes trapping at the ESP}.
Note also, that, while $H_*$ is determined by the location of the ESP in field 
space (by the vacuum density $V_0$), the values of $|m_\phi|$ and 
$|V^{\textrm{\tiny{(3)}}}_0|$ 
are due to the deformation of the scalar potential at the vicinity of the ESP 
which is not directly related to $V_0$. Hence, the requirement that the latter 
are smaller than $H_*$ does not necessarily imply fine-tunning.

\section{After the end of inflation}

After inflation, the field rolls away from the ESP. Soon the 
influence of the ESP on the scalar potential diminishes and 
$V(\phi)\approx V_{\rm np}(\phi)$.
The steepness of $V_{\rm np}$ results in the kinetic domination of the modulus
density. As a result a period of kination occurs, during which the field
equation is: $\ddot\phi+3H\dot\phi\simeq 0$. Hence, the density
of the Universe scales as $\rho\simeq\frac{1}{2}\dot\phi^2\propto a^{-6}$ 
\cite{kination}. During kination, the scalar field is oblivious 
of the particular form of the potential.

Kination is terminated when the density of the decay products of a curvaton 
field dominates the kinetic density of the modulus \cite{curvreh}. 
Thus, the end of kination  corresponds to reheating, with 
reheating temperature $T_{\rm reh}\sim\sqrt{H_{\rm reh}m_P}$, where
$H_{\rm reh}$ is the Hubble parameter at reheating.

\begin{figure}
\hspace{2cm}
\psfig{file=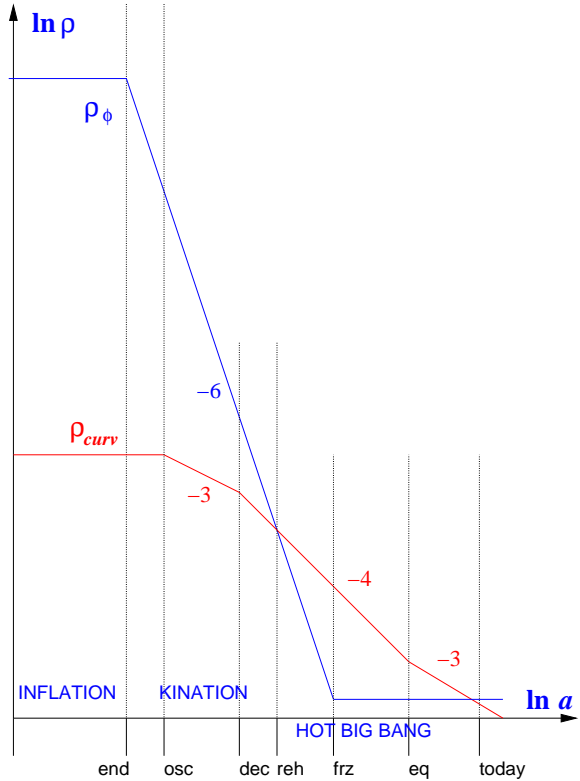,width=3in}
\caption{%
Illustration of the evolution of the modulus density $\rho_\phi$ and the 
density of the curvaton and its decay products $\rho_{\rm curv}\,$ with respect
to the scale factor of the Universe $a$. In inflation, $\rho_{\rm curv}\,$ 
is subdominant and remains constant until, after the end of inflation (denoted 
by `end') the curvaton begins oscillating (at time denoted by `osc'). During 
the oscillations, $\rho_{\rm curv}\,$ scales as pressureless matter. Sometime 
afterwards (denoted by `dec') the curvaton decays into the thermal bath of the 
HBB. This thermal bath dominates the Universe at reheating (denoted by `reh') 
soon after which the modulus freezes (at time denoted by `frz') assuming 
constant potential density comparable to the density today.}
\label{kin}
\end{figure}

After the onset of the HBB, the rolling scalar field is subject to cosmological
friction \cite{eta,cosmofric}, which asymptotically freezes the field at the 
value $\phi_F/m_P\simeq\frac{1}{\sqrt 6}\ln(V_0/T_{\rm reh}^4)$. Note that 
this value depends on $T_{\rm reh}$ which, in turn, is determined by curvaton 
physics. The modulus remains frozen until the present when it 
plays the role of quintessence. This guarantees that there is no dangerous 
variation of fundamental constants during the HBB. 
The evolution of the modulus until today is depicted in Fig.~\ref{kin}.

\section{Quintessence}

Since $\sigma_F\equiv\sigma(\phi_F)\sim(V_0/T_{\rm reh}^4)^{1/3}>1$,
the modulus rolls to large values before freezing.
At such values we can assume that the scalar potential is 
\begin{equation}
\label{exps}
V(\sigma)\simeq\frac{C_n}{\sigma^n}
\quad\Rightarrow\quad 
V(\phi)\simeq C_ne^{-b\phi/m_P}\,,
\end{equation}
where $C_n$ is a density scale and $b=\sqrt{\frac{2}{3}}\,\,n$. 
The above is a typical uplift potential introduced by flux compactifications
as discussed below.

If the modulus is to account for the required dark energy, it must
satisfy the coincidence requirement:
$V(\sigma_F)\simeq\Omega_\Lambda\rho_0$, where $\Omega_\Lambda\simeq 0.73$ is
the dark energy density parameter and $\rho_0$ is the critical density at 
present. Hence, 
\begin{equation}
\label{Treh}
T_{\rm reh}\sim V_0^{1/4}
\left(\rho_0/C_n\right)^{\sqrt{3/8n^2}}.
\end{equation}
Thus, the density scale $C_n$ is determined by  $T_{\rm reh}$ which, in turn, 
is determined by curvaton physics. An upper bound on $C_n$ is obtained by
demanding that reheating occurs before big bang nucleosynthesis (BBN):
\begin{equation}
\label{Cbound}
C_n\lsim\rho_0\left(V_0^{1/4}/T_{\rm BBN}\right)^{2n\sqrt{2/3}},
\end{equation}
where $T_{\rm BBN}\sim 1$~MeV is the temperature at BBN.

The scalar potential in Eq.~(\ref{exps}) may have a multitude of origins.
For example, using the volume modulus, we may consider a stack of 
$\overline{D3}-$branes located at the tip of a Klebanov-Strassler throat.
The uplift potential is
 \cite{Giddings:2001yu}
\begin{equation}
\delta V\sim\exp(-8\pi K/3Mg_s)\,m_P^4/\sigma^2
\equiv C_2/\sigma^2\,,
\end{equation}
where $M$ and $K$, in the warp factor, are the units of RR and NS three-form 
fluxes. To satisfy Eq.~(\ref{Cbound}) we must have $C_2^{1/4}\lsim10^{-20}m_P$.
This can be attained by choosing the ratio of fluxes as $K/Mg_s\gsim22$. 
Taking $g_s=0.1$, only twice as many units of $K$ flux as those 
of $M$ flux are needed.

It is also possible to consider fluxes of gauge fields on
$D7-$branes \cite{Burgess:2003ic}. 
In this case, the scalar potential obtains a contribution
\begin{equation}
\delta V\sim 2\pi E^2/\sigma^3
\equiv C_3/\sigma^3\,,
\end{equation}
where $E$ depends on the strength of the gauge fields considered.
The constraint in Eq.~(\ref{Cbound}) requires now
$C_3^{1/4}\lsim 10^{-15}m_P\sim 1$~TeV.

The future of the modulus after unfreezing depends on the steepness of the 
scalar potential, or equivalently the value of $b$ in Eq.~(\ref{exps}). 

\noindent{$\bullet$}
For $b\leq\sqrt 2$, the modulus dominates the Universe for ever, leading to 
eternal acceleration. This results in future horizons, which pose a problem
for the formulation of the S-matrix in string theory \cite{S}.

\noindent{$\bullet$}
For $\sqrt 2< b\leq\sqrt 3$, the modulus dominates the Universe but results 
only in a brief accelerated expansion period. 
Such is the fate of the $n=2$ case. 

\noindent{$\bullet$}
For $\sqrt 3<b\leq\sqrt 6$, the modulus does not dominate the Universe, albeit
causing a brief period of accelerated expansion. Afterwards the modulus
density remains at a constant ratio with the background matter density.
This is the fate of the $n=3$ case.

\noindent{$\bullet$}
For $b>\sqrt 6$, the modulus does not cause any accelerated 
expansion and so cannot be used as quintessence. After unfreezing, the
modulus rolls fast down the quintessential tail of the scalar potential with
its density approaching asymptotically kinetic domination (and subsequently
freezing at a value larger than $\sigma_F$ \cite{eta}). 
This case corresponds to $n>3$.


The brief acceleration period caused by the unfreezing modulus is due to the
fact that the modulus oscillates around an attractor solution \cite{oscil},
which in itself does not result to acceleration \cite{jose}. 
In Ref.~\cite{cline} it was claimed that brief acceleration occurs if 
$\sqrt 2<b\leq 2\sqrt 6$, which corresponds to the range $\sqrt 3<n\leq 6$. 
More recent studies, however, have reduced this range. Brief acceleration in 
the range $\sqrt 2<b\leq\sqrt 3$ has been confirmed by Ref.~\cite{rosen}. This 
includes the $n=2$ case which corresponds to the most popular uplift potential.
The range for $b$ was expanded further in Ref.~\cite{blais}, where it is shown
that brief acceleration can explain the observations at least 
up to \mbox{$b\simeq\frac{3}{4}\sqrt 6\approx 1.837$}. Since the 
data are interpreted using a number of priors, we believe that the $n=3$
case is still marginally acceptable. 

\section{Conclusions}

Quintessential inflation is possible to achieve in string theory 
with flux compactifications using as inflaton a string modulus which rolls 
down its runaway potential.
Inflation is due to the presence of an enhanced symmetry point (ESP),
which traps the modulus and creates a locally flat region over which the 
modulus can slow-roll. There is ample parameter space
for successful inflation provided: $m_\phi,|V^{\textrm{\tiny{(3)}}}_0|\ll H_*$.
Trapping assimilates a multitude of initial conditions provided overshooting 
the ESP is avoided. 

After inflation, the modulus becomes (again) kinetically dominated causing
a period of kination.
Reheating is due to the Universe domination by the decay products of a 
curvaton field, which also accounts for the correct amplitude of the curvature 
perturbations in the Universe.
During the Hot Big Bang, the modulus freezes and remains constant until the 
present.

At the frozen value, the potential is dominated by an uplift term of 
exponential form. This residual potential density begins to dominate today, 
when the modulus unfreezes, leading to a brief period of late inflation.

Curvaton physics fixes the reheating temperature and determines the
value of the frozen modulus $\sigma_F$ and the density scale $C_n$
in the uplift potential. Coincidence and BBN constrains on $C_n$
allow realistic values much less fine-tunned than 
the cosmological constant $\Lambda$ in the $\Lambda$CDM model.

\section*{Acknowledgements}

This work was done with Juan C. Bueno S\'{a}nchez and was supported 
by the E.U. Marie Curie Research and Training Network 
{\sl ``UniverseNet"} (MRTN-CT-2006-035863) and by PPARC (PP/D000394/1).
I am grateful to the Royal Society for supporting my participation to the 
IDM2006 conference.

\end{document}